\documentclass[preprint,showpacs,preprintnumbers,amsmath,amssymb, aps, pra, nofootinbib]{revtex4-2}
\eprint
\usepackage{lipsum}
\usepackage[symbol]{footmisc}
\usepackage{graphicx}
\usepackage{amsmath}
\usepackage{tabularx}
\usepackage{mathtools}

\DeclarePairedDelimiterX\braket[2]{\langle}{\rangle}{#1 \delimsize\vert #2}
\usepackage{bm}
\usepackage{color}
\usepackage{esvect}
\usepackage{gensymb}
\usepackage{parskip}

\usepackage{ulem}

\usepackage{comment}
\usepackage[colorlinks=true, citecolor=blue, linkcolor=blue, urlcolor=blue]{hyperref}
\usepackage{tikz,lipsum,lmodern}
\usepackage[most]{tcolorbox}

\begin{document}

\title{Nonresonant nonlinear magnonics in an antiferromagnet}

\author{Gu-Feng Zhang$^{1,11}$} 
\author{Sheikh Rubaiat Ul Haque$^{1,2,3,11}$} \email{rubaiath@stanford.edu}
\author{Kelson J. Kaj$^{1,4,11}$} 
\author{Xiang Chen$^5$}
\author{Urban F. P. Seifert$^{6,7}$}
\author{Jingdi Zhang$^{1,8}$}
\author{Kevin A. Cremin$^1$}
\author{Leon Balents$^{6,9,10}$}
\author{Stephen D. Wilson$^5$}
\author{Richard D. Averitt$^1$} \email{raveritt@ucsd.edu}

\affiliation{
$^1$Department of Physics, University of California San Diego, La Jolla, CA 92093, USA\\
$^2$Department of Applied Physics, Stanford University, Stanford, CA 94305, USA\\
$^3$Stanford Institute for Materials and Energy Sciences, SLAC National Accelerator Laboratory, Menlo Park, CA 94025, USA\\
$^4$Max Planck Institute for the Structure and Dynamics of Matter, Luruper Chausse 149, 22761 Hamburg, Germany\\
$^5$Materials Department, University of California Santa Barbara, CA 93106, USA\\
$^6$Kavli Institute for Theoretical Physics, University of California Santa Barbara, CA 93106, USA\\
$^7$Institute for Theoretical Physics, University of Cologne, 50937 Cologne, Germany\\
$^8$Department of Physics, The Hong Kong University of Science and Technology, Clear Water Bay, Kowloon, Hong Kong SAR, China\\
$^9$Canadian Institute for Advanced Research, Toronto, ON M5G 1M1, Canada\\
$^{10}$French-American Center for Theoretical Science, CNRS, KITP, Santa Barbara, CA 93106, USA\\
$^{11}$These authors contributed equally to the work
}
 
\maketitle

\noindent \textbf{Antiferromagnets exhibit rapid spin dynamics in a net zero magnetic background which enables novel spintronic applications and interrogation of many-body quantum phenomena. The layered antiferromagnet Sr$_2$IrO$_4$ hosts an exotic spin one-half Mott insulating state with an electronic gap arising from on-site Coulomb repulsion and strong spin-orbit coupling. This makes Sr$_2$IrO$_4$ an interesting candidate to interrogate dynamical attributes of the magnetic order using ultrafast laser pulses. We investigate the magnetization dynamics of Sr$_2$IrO$_4$ following circularly-polarized photoexcitation with below-gap mid-infrared (mid-IR -- 9 $\mu m$) and above-gap near-infrared (near-IR -- 1.3 $\mu m$) pulses. In both cases, we observe excitation of a zone-center coherent magnon mode featuring a 0.5 THz oscillation in the pump-induced Kerr-rotation signal. However, only below-gap excitation exhibits a helicity dependent response and linear (quadratic) scaling of the  coherent magnon amplitude with excitation fluence (electric field). Moreover, below-gap excitation has a magnon generation efficiency that is at least two orders of magnitude greater in comparison to above-gap excitation. Our analysis indicates that the helicity dependence and enhanced generation efficiency arises from a unique one-photon two-magnon coupling mechanism for magnon generation. Thus,  preferential spin-photon coupling without photoexcitation of electrons permits extremely efficient magnon generation. Our results reveal new possibilities for ultrafast control of antiferromagnets.}


In strongly correlated materials, the interplay between different degrees of freedom govern  emergence and functionality \cite{Zhang2014}\cite{Basov2017}\cite{Basov2011}. Of particular interest is the study of exotic magnetic phases. Magnetically ordered systems are amenable to  nonequilibrium optical manipulation \cite{Disa2020}\cite{Disa2023}\cite{Zhang2014}\cite{Basov2017}\cite{DeLaToore2021}\cite{Kampfrath2013}\cite{Kirilyuk2010}\cite{Ashoka2023}\cite{Stoica2022}\cite{Ron2020}\cite{Lovinger2020}, and advances in ultrafast laser technology have enabled experiments to manipulate and control collective excitations (magnons) on pico- or femtosecond timescales \cite{Dean2016}\cite{Mazzone2021}\cite{Matthiesen2023}\cite{Afanasiev2021}\cite{Afanasiev2019}\cite{Blank2023}\cite{Mashkovich2021}\cite{Grishunin2021}\cite{Mikhaylovskiy2020}\cite{Kampfrath2011}\cite{Zhang20241}\cite{Zhang20242}\cite{Huang2024}. Recently, antiferromagnets (AFM) have drawn considerable interest since the magnetic dynamics are considerably faster than ferromagnets (FM) \cite{Kimel2009}\cite{Nemec2018} due to their zero net magnetization. This remarkable property paves the way towards ultrafast magnetic control with intense laser pulses \cite{Kampfrath2011} and spintronic devices. In general, lightwave-control of magnetism can be achieved in several ways. This includes direct coupling between the magnetic field of light and spins \cite{Kampfrath2011}\cite{Baierl2016}, modification of exchange interactions \cite{Mikhaylovskiy2015}\cite{Mentink2015}, intermediate excitation such as phonons \cite{Mashkovich2021}\cite{Nova2017}, or magneto-optical effects, including the inverse Faraday effect and the inverse Cotton-Mouton effect \cite{Kimel2005}\cite{Mikhaylovskiy2012}\cite{Satoh2010}\cite{Tzschaschel2017}\cite{Kalashnikova2007}\cite{Kalashnikova2008}\cite{Afanasiev2014}. Hence, harnessing the potential of coupling magnetism with optical stimuli may provide insights on many long-standing puzzles in strongly correlated systems and facilitate next-generation information technologies.

 The ultrafast manipulation of magnetism in correlated materials can take advantage of the magnetic phases that emerge from the competition between the on-site Coulomb interaction and spin-orbit coupling. Iridium oxides (iridates) have attracted attention as $5d$ electron systems where the competing Coulomb interaction and spin-orbit coupling are of comparable strength \cite{Rau2016}\cite{Witczak2014}\cite{Cao2018}. This may also permit various intriguing quantum phases, including topological Mott insulators, spin liquids, and quadrupolar order \cite{Witczak2014}\cite{Pesin2010}. Along these lines, Sr$_2$IrO$_4$, a spin-orbit-coupled Heisenberg AFM Mott insulator with a meV-scale in-plane spin gap \cite{Gim2016}\cite{Gretarsson2017}\cite{Porras2019}\cite{Liu2019}, has emerged as a rich platform to explore photon-magnon interactions and ultrafast manipulation of magnetic order.
 
\setcounter{figure}{0}
\renewcommand{\figurename}{\textbf{ Fig.}}
\renewcommand{\thefigure}{\arabic{figure}}

\begin{figure}
    \centering
    \includegraphics[width=140 mm]{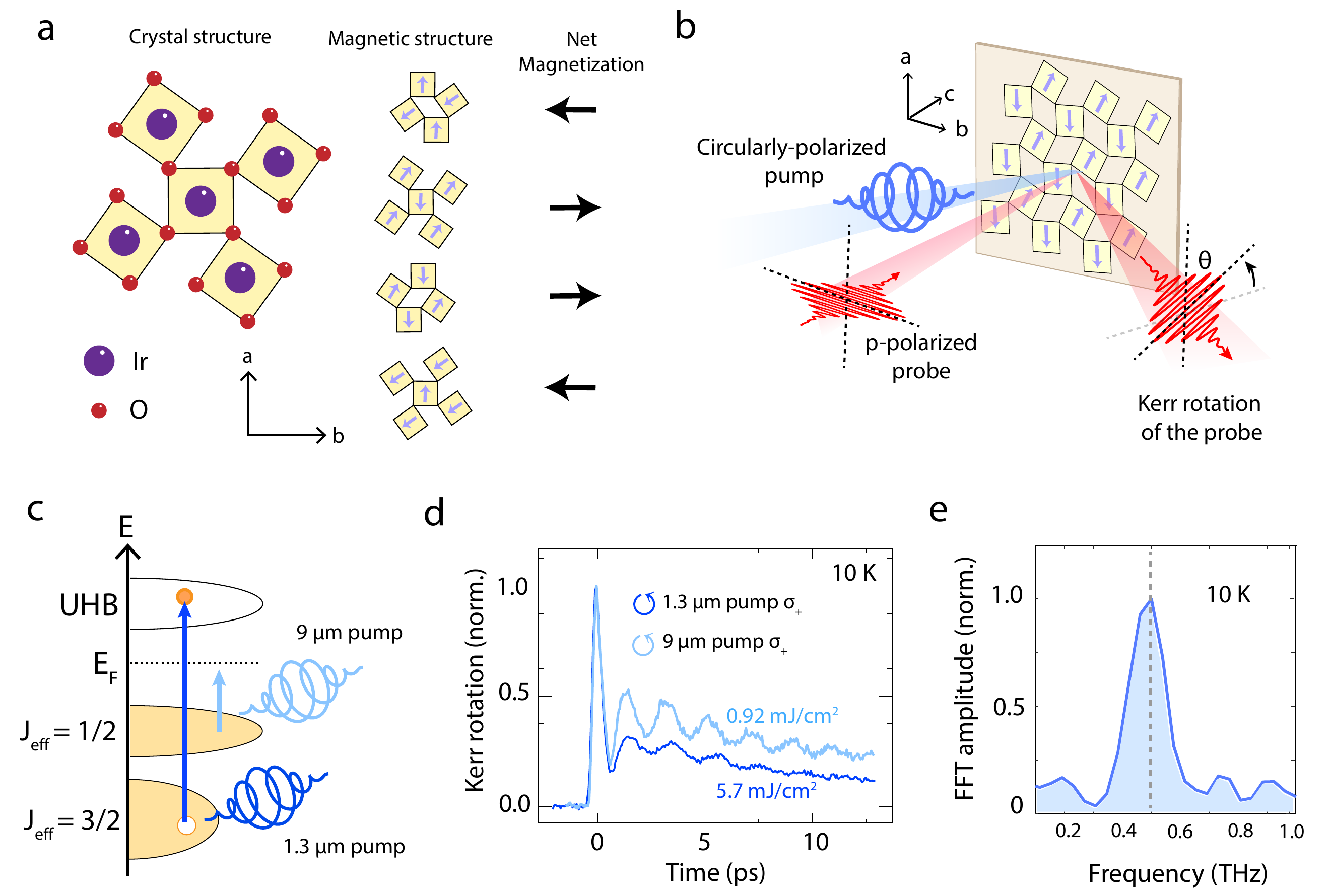}
    \caption{\textbf{Introduction to Sr$_2$IrO$_4$ and experimental configuration. a}, Staggered Ir-O planes of Sr$_2$IrO$_4$ displaying AFM ordering along the c-axis. \textbf{b}, Experimental scheme: circularly-polarized pump pulses at 9 $\rm \mu m$ or 1.3 $\rm \mu m$ (blue) are incident on ab-plane oriented Sr$_2$IrO$_4$. A p-polarized 800 nm probe pulse (red) is reflected from the sample, and the time-resolved magneto-optical Kerr effect is measured (i.e., polarization rotation $\theta$ of the probe pulses). \textbf{c}, Schematic density of state (DOS) of Sr$_2$IrO$_4$. The 9 $\rm \mu m$ pump is below the electronic band gap, and above the highest optical phonon and is thus not resonant to any dipole-active excitation. In contrast, the 1.3 $ \mu m$ pump resonantly excites the electrons from the $J_{eff}$$=3/2$ band to the upper Hubbard band (UHB). \textbf{d}, Time-resolved Kerr rotation  at 10 K from circularly-polarized 1.3 $\rm \mu m$ (dark blue) and 9 $\rm \mu m$ (light blue) excitation with similar helicity (left circularly-polarized light, $\circlearrowleft$ : $ \sigma_+$). Note that, for magnon oscillations with comparable amplitude shown here, the pump fluences for 9 $\rm \mu m$ and 1.3 $\rm \mu m$ pump are 0.92 $\rm mJ/cm^2$ and  5.7 $\rm mJ/cm^{2}$, respectively. \textbf{e}, Fourier transform of the Kerr signal for the 9 $\rm \mu m$ excitation, revealing a $B_{2g}$ magnon at 0.5 THz.}
    \label{fig:1_schematic}
\end{figure}

Sr$_2$IrO$_4$ exhibits a layered quasi-2D structure with $J_{eff}=1/2$ pseudospins at each IrO$_6$ octahedra that order antiferromagnetically within a given Ir-O plane below $T_N$ $\sim$ 230 K \cite{Ye2013}. In the AFM phase, the system forms a staggered AFM order with a $-++-$ pattern along the c-axis (Fig. \ref{fig:1_schematic}a, net magnetic moment shown in black arrows) which can transition to a weakly FM state ($++++$) via an external in-plane magnetic field of $H_c$ $\sim$ 0.2 T \cite{Porras2019}\cite{Kim2009}\cite{Cao1998}. The slight canting of the spins results in a magneto-optic Kerr effect (MOKE) which is key to tracking the photoinduced magnetic dynamics. Crystal field splitting of $5d$ electron states leads to a partially-filled $t_{2g}$ manifold which is further split into a filled $J_{eff}=3/2$ band and a half-filled $J_{eff}=1/2$ band by strong spin-orbit coupling. Finally, Coulomb repulsion splits the $J_{eff}=1/2$ band into upper and lower Hubbard bands, opening an electronic gap of $\sim0.5$ eV (Fig. \ref{fig:1_schematic}b-c) \cite{Kim2008}\cite{Bertinshaw2018}. 

A thorough study of the magnetic interactions in Sr$_2$IrO$_4$ warrants experiments with access to the low-energy magnons. In this work, we investigate the ultrafast magnonic dynamics in Sr$_2$IrO$_4$ utilizing mid- and near-infrared pump - time-resolved MOKE probe experiment (see Fig. \ref{fig:1_schematic}b). By pumping Sr$_2$IrO$_4$ with circularly-polarized light at different wavelengths, we measure the dynamical magnetic response to excitations both above and below the electronic gap. We emphasize that strong coherent magnon oscillations can be launched by below-gap pumping at remarkably lower excitation fluences compared to above-gap pumping, as shown in Fig. \ref{extfig:1_schematic}d-e. These findings demonstrate efficient magnon generation and control using nonresonant laser fields. 

\section*{Photoinduced coherent magnon dynamics}
\label{sec:Results}

\begin{figure*}
    \centering
    \includegraphics[width=160 mm]{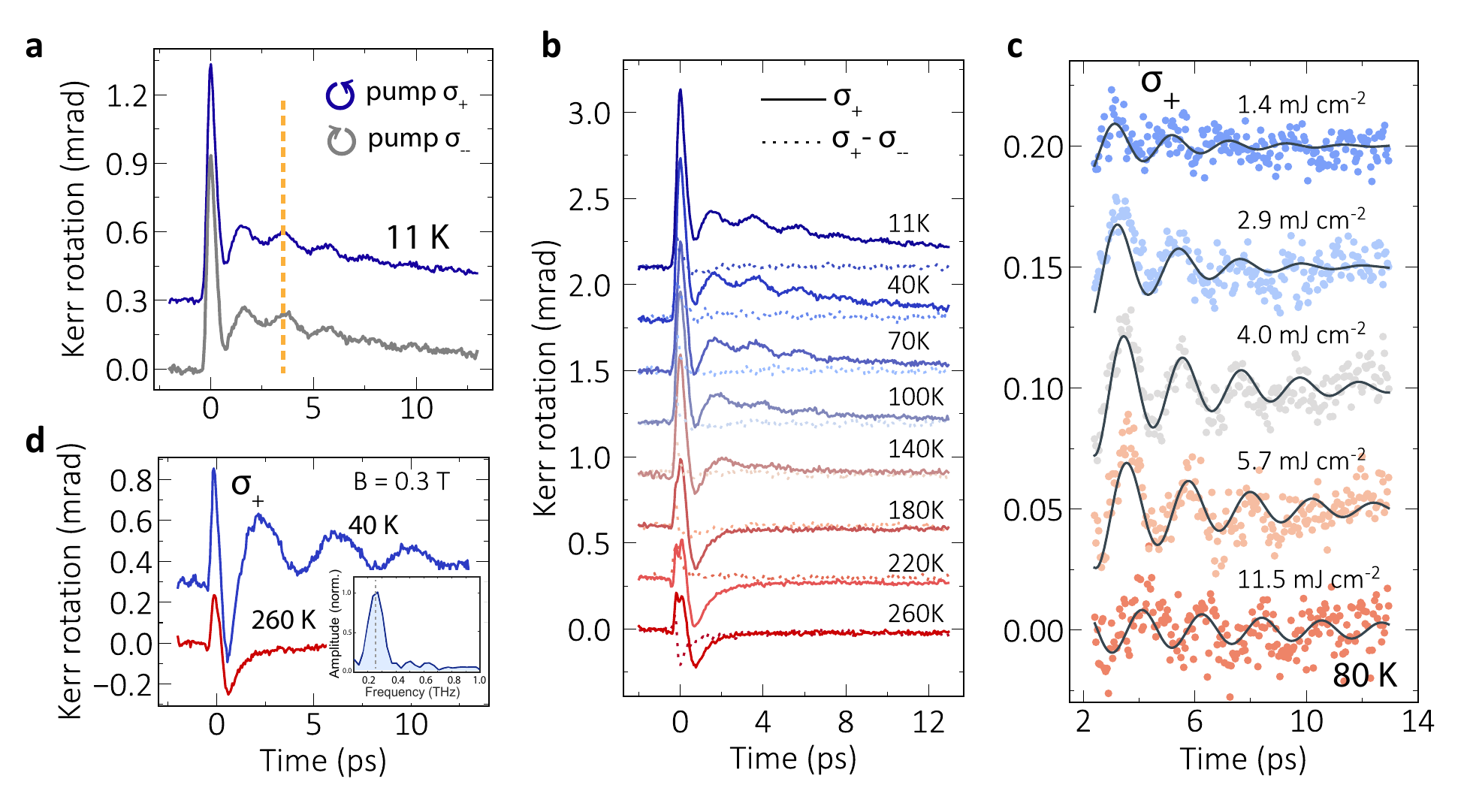}
    \caption{\textbf{Kerr rotation dynamics for 1.3 $\rm \mu m$ circularly-polarized pump excitation.} The pump fluence is 5.7 $\rm mJ/cm^2$ for \textbf{a}, \textbf{b}, and \textbf{c}. \textbf{a}, Kerr rotation dynamics upon $\sigma_+$ ($\circlearrowleft$, blue) and $\sigma_-$ ($\circlearrowright$, gray) excitation at 11 K. The orange vertical dashed line is a guide to the eye, showing that magnon oscillations are in-phase for both helicities. The $\sigma_-$ pump signal is shifted vertically for clarity. \textbf{b}, Temperature-dependent Kerr rotation dynamics following $\sigma_+$ pump (solid lines). The differential time traces between $\sigma_+$ and $\sigma_-$ pumping ($\sigma_+ - \sigma_-$, dashed lines) are featureless at all temperatures highlighting the lack of helicity dependence. \textbf{c}, Fluence-dependent $\sigma_+$ pump magnon dynamics. Solid lines are fits with a damped oscillator model with the exponential contribution subtracted. \textbf{d}, $\sigma_+$ pump dynamics at 40 K (blue) and 260 K (red) with a 0.3 T magnetic field applied along the a-axis, displaying a 0.26 THz resonance (inset).}
    \label{fig:2_1p3um}
\end{figure*}

We performed time-resolved magneto-optical Kerr effect (MOKE) measurements on Sr$_2$IrO$_4$ following excitation with circularly-polarized above-gap near-IR 1.3 $\rm \mu m$ and below-gap mid-IR 9 $\rm \mu m$ pump pulses. Fig. \ref{fig:2_1p3um} depicts the Kerr rotation dynamics for circularly-polarized 1.3 $\rm \mu m$ (0.95 eV) pumping. The Kerr rotation dynamics at 11 K upon both left ($\circlearrowleft$ : $\sigma_+$) and right ($\circlearrowright$ : $\sigma_-$) circularly-polarized excitation are shown in Fig. \ref{fig:2_1p3um}a. For both pump helicities, long-lived coherent oscillations in the Kerr rotation signal are recorded. The frequency of the oscillation is $\sim 0.5$ THz (2 meV), consistent with the $B_{2g}$ one-magnon resonance as reported in Raman measurements \cite{Gim2016}\cite{Gretarsson2017}. The magnon oscillations are in-phase for both $\sigma_+$ and $\sigma_-$ pump helicities, indicated by the orange vertical dashed line in Fig. \ref{fig:2_1p3um}a. The temperature-dependent photoinduced MOKE signal for $\sigma_+$ pump (solid lines) as well as the differential time trace signal between $\sigma_+$ and $\sigma_-$ photoexcitation ($\sigma_+ - \sigma_-$, dashed lines) are shown in Fig. \ref{fig:2_1p3um}b, respectively.  The coherent magnon oscillation of the $\sigma_+$ pump persists up to 180 K and redshifts with temperature, consistent with the fact that the 0.5 THz $B_{2g}$ magnon only appears in the AFM phase below the $T_N$ $\sim$ 230 K. In contrast, no oscillations are observed for the $\sigma_+ - \sigma_-$ dynamics at any temperature, clearly showing the lack of a pump helicity dependence (in stark contrast with below-gap excitation as described below).  Fig. \ref{fig:2_1p3um}c plots the fluence-dependent $\sigma_+$ pump-induced dynamics at 80 K. The coherent magnon oscillation persists up to the highest used fluence of 11.5 $\rm mJ/cm^2$, which is sufficient to destroy the 3D magnetic order, but not the 2D magnetic correlations \cite{Dean2016}. This is suggestive of the in-plane 2D nature of the $B_{2g}$ magnon. Furthermore, applying an in-plane external magnetic field of 0.3 T along the a-axis of the sample (Fig. \ref{fig:2_1p3um}d) results in a redshift of the magnon oscillation to $\sim0.26$ THz. A similar redshift of the $B_{2g}$ magnon with applied magnetic field has been observed in Raman measurements \cite{Gim2016} as Sr$_2$IrO$_4$ undergoes an AFM ($-++-$) to weakly FM ($++++$) phase transition. This confirms that the observed coherent oscillation is the $B_{2g}$ magnon.

\begin{figure*}
    \centering
    \includegraphics[width=120 mm]{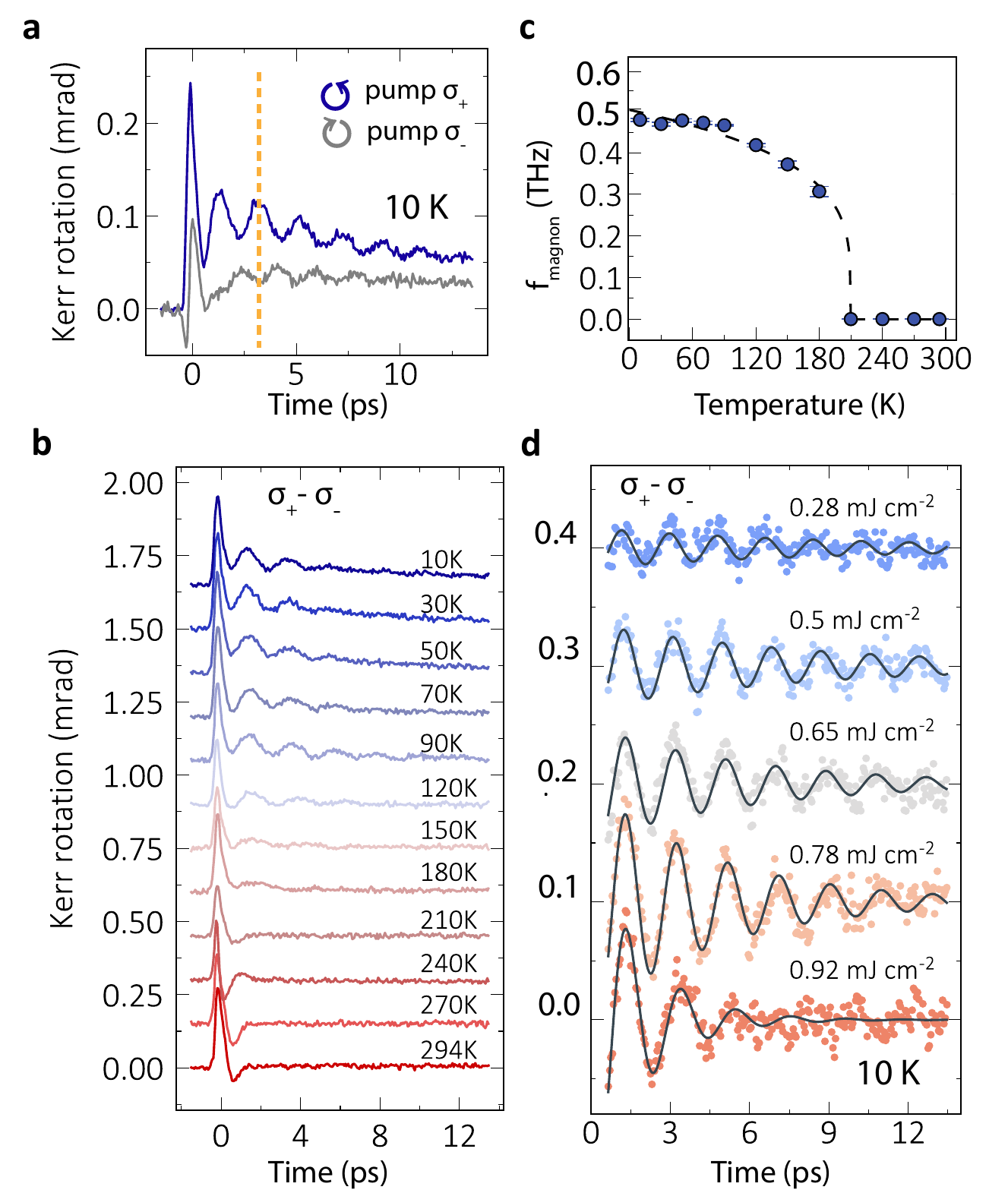}
    \caption{\textbf{Kerr rotation dynamics for 9 $\rm \mu m$ circularly-polarized excitation.} For \textbf{a} and \textbf{b}, the pump fluence is 0.92 $\rm mJ/cm^2$. \textbf{a}, Kerr rotation dynamics with opposite pump helicities ($\sigma_+$ and $\sigma_-$) at 10 K. The orange vertical dashed line is a guide to the eye, showing that the the magnon oscillations are out-of-phase. \textbf{b}, Dynamics of $\sigma_+ - \sigma_-$ at different temperatures. \textbf{c}, Temperature dependence of the magnon oscillation frequency $f_{magnon}$. The dashed line fits the magnon frequency to the power law scaling $\left | (1-T/T_N)^{2\beta} \right |$ with $\beta = 0.134$, a value close to $ \frac{1}{8}$, as expected for a 2D Heisenberg model \cite{Binder76}. Error bars represent the s.e.m. from four independent measurements. \textbf{d}, Fluence-dependence of $\sigma_+ - \sigma_-$ at 10 K. The solid lines denote fits with a damped oscillator model with the exponential contribution subtracted. }
    \label{fig:3_9um}
\end{figure*}

We now describe the magnon dynamics for below-gap 9 $\rm \mu m$ (138 meV) circularly-polarized pump excitation. This photon energy is well below the electronic gap ($\sim 0.5$ eV) and greater than the highest energy IR-active optical phonon ($\sim 82$ meV) \cite{Moon2009} (Fig. \ref{fig:1_schematic}c), and is thus nonresonant with any dipole-allowed phonon or electronic transitions. Fig. \ref{fig:3_9um}a displays the Kerr rotation dynamics for circularly-polarized pumping with opposite helicities ($\sigma_+$ and $\sigma_-$) at 10 K. The 0.5 THz coherent $B_{2g}$ magnon is generated for both helicities. However, the corresponding magnon oscillations are 180$\degree$ out-of-phase, as shown by the orange vertical dashed line in Fig. \ref{fig:3_9um}a. This contrasts with the 1.3 $\rm \mu m$ pumping described above. Subtracting the $\sigma_-$ pump response from that of $\sigma_+$ yields the purely magnetic response, $\sigma_+ - \sigma_-$. The temperature dependence of the $\sigma_+ - \sigma_-$ dynamics are plotted in Fig. \ref{fig:3_9um}b. The coherent magnon oscillation is visible up to $\sim210$ K, disappearing at higher temperatures. The temperature-dependent magnon frequency is plotted in Fig. \ref{fig:3_9um}c. The order parameter-like temperature dependence of the magnon frequency  demonstrates that the magnon scales with the AFM ordering.

Figure \ref{fig:3_9um}d shows the fluence dependent $\sigma_+ - \sigma_-$ dynamics at 10 K, revealing that the coherent magnon is observed over the entire range of fluences utilized in these experiments. Notably, for below-gap 9 $\rm \mu m$ pumping, magnon oscillations of comparable amplitude are generated with fluences at least one order of magnitude smaller than for above-gap pumping. This is underscored in Fig. \ref{fig:3_9um}d where the 9 $\rm \mu m$ pump with a modest fluence of 0.5 $\rm mJ/cm^2$  launches stronger magnon oscillation than 1.3 $\rm \mu m$ pumping at much higher fluences (Fig. \ref{fig:2_1p3um}c). Surprisingly, the pump penetration depth for 9 $\rm \mu m$ (1217 nm) excitation is one order of magnitude larger than that of 1.3 $\rm \mu m$ pump (141 nm). Dividing the pump fluence by its penetration depth provides an estimate of the energy density in units of $\rm mJ/cm^3$ within the photoexcited volume. The pronounced disparity in fluence and penetration depth between the two pumping conditions emphasizes that the below-gap 9 $\rm \mu m$ pump generates strong magnon oscillations, despite having a pump energy density two orders of magnitude lower than the above-gap 1.3 $\rm \mu m$ pump.

\section*{Microscopic mechanisms for below- and above-gap photoexcitation}

\begin{figure*}[h!]
    \centering
    \includegraphics[width=130 mm]{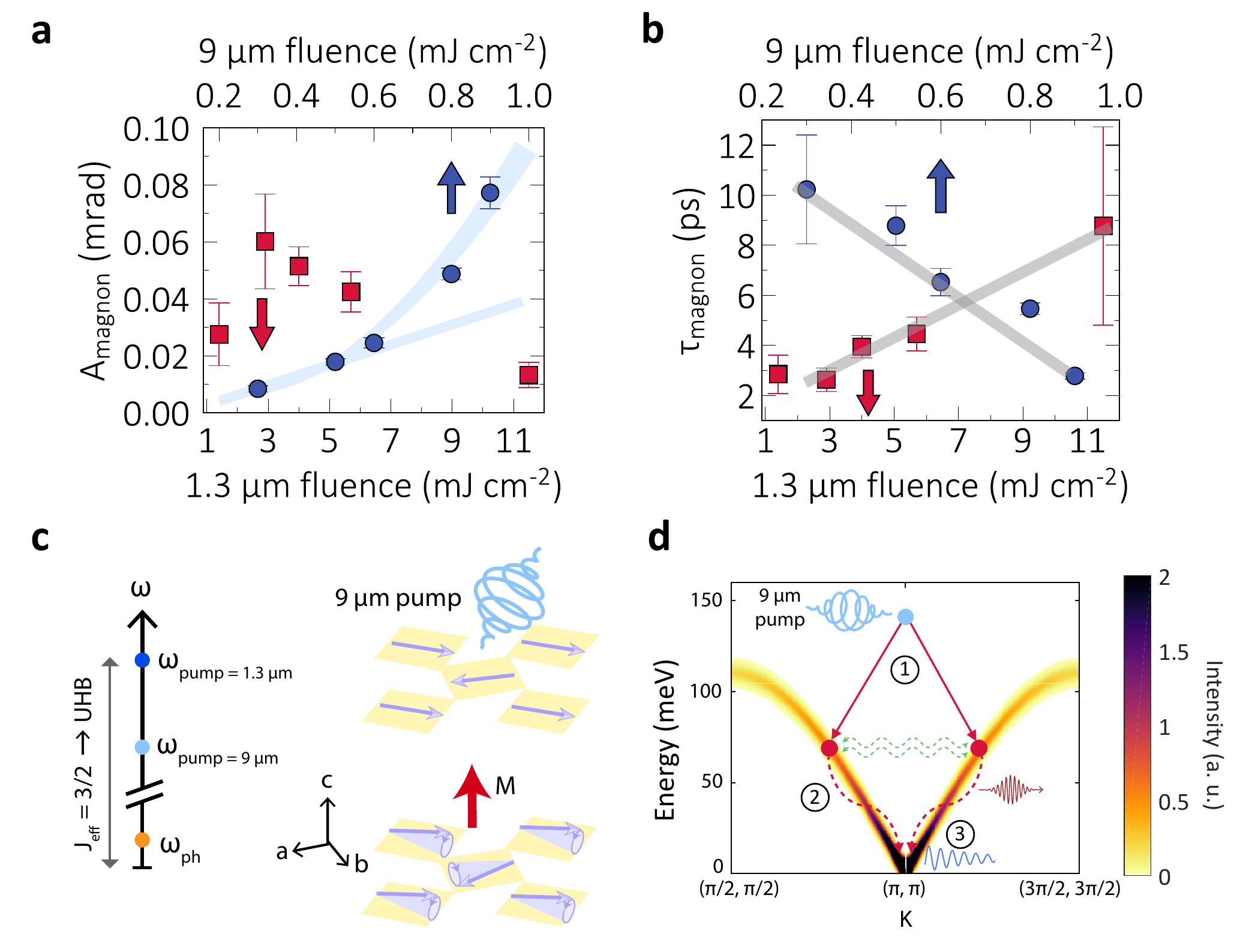}
    \caption{\textbf{Fluence dependence and schematic of the magnon generation mechanism. a}, Magnon oscillation amplitude A$_{magnon}$ shows opposite fluence-dependent trends for 9 $\rm \mu m$ (solid blue circles) and 1.3 $\rm \mu m$ pumping (solid red squares). A$_{magnon}$ exhibits linear scaling at lower fluences and quadratic scaling at higher fluences, as demonstrated by the fit within the shaded region for 9 $\rm \mu m$ pump. However, for 1.3 $\rm \mu m$ pump, A$_{magnon}$ decreases with higher fluence ($>$ 3 $\rm mJ/cm^2$). Error bars represent the s.e.m. from four independent measurements. \textbf{b}, Damping time constant $\tau_{magnon}$ showing decrease for the 9 $\rm \mu m$ pump fluence while increase for the 1.3 $\rm \mu m$ pump fluence. Gray lines are guide to the eye. \textbf{c}, The energy diagram showing 9 $\rm \mu m$ pump below the electronic gap but above the highest energy optical phonon $\omega_{ph}$. Depiction of photoinduced magnon oscillations and out-of-plane magnetization (red arrow). \textbf{d}, Schematic of the 9 $\rm \mu m$ pump magnon generation mechanism near ($\pi$, $\pi$) in the Sr$_2$IrO$_4$ spin spectrum in three steps: (1) the 9 $\rm \mu m$ (138 meV) photon couples to a pair of (virtual) magnons of equal and opposite momenta, (2) these magnons decay to the low-energy 2 meV magnon at ($\pi, \pi$) and an outgoing photon via three-magnon -- one-photon scattering processes (green wavy lines), and (3) the 2 meV $B_{2g}$ magnon oscillation probed by MOKE. }
    \label{fig:4_theory}
\end{figure*}

The observed long-lived $B_{2g}$ magnon oscillation at $\sim0.5$ THz is indicative of a small magnon gap whereas a 2D square lattice Heisenberg model is known to involve a gapless Goldstone mode. However, incorporating a small exchange anisotropy $\Gamma$ into the Heisenberg model opens a spin gap. The interaction can be written as $\sum_{\left \langle i,j \right \rangle}\Gamma(S_i^xS_j^y+S_i^yS_j^x) $. Here, $\Gamma\sim3$ $\mu$eV and it stems from strong spin-orbit coupling and Jahn-Teller distortions \cite{Porras2019}\cite{Liu2019}. Meticulous efforts to investigate the spin gap have been made using Raman scattering, resonant inelastic X-ray scattering (RIXS) and inelastic neutron scattering (INS) experiments \cite{Ye2013}\cite{Gretarsson2016}\cite{Pincini2017}\cite{Kim2014}\cite{Liu2016}\cite{Cao2017}\cite{Cao2016}\cite{Porras2019}, and recently a 2 meV spin gap at ($\pi$, $\pi$) point in reciprocal space was reported via RIXS \cite{Porras2019}. Hence, the physics of Sr$_2$IrO$_4$ can be approximately explained by an XY model with a small in-plane two-fold anisotropy that gives rise to a weakly-gapped magnon mode. Further, the coherent magnon we observe corresponds to precession at ($\pi$, $\pi$). We note that the $(\pi, \pi)$ point is expressed using a smaller unit cell notation used in literature \cite{Porras2019}, and it maps onto the ($2\pi$, 0) point of the true chemical unit cell, carrying the same crystal momentum as the $\Gamma$ point.

To gain insight into the microscopic mechanism of the photoinduced magnon generation dynamics, we first showcase the key differences between the 1.3 $\rm \mu m$ and 9 $\rm \mu m$ excitation data. Looking at the dielectric permittivity tensor $\varepsilon_{ij}=\varepsilon_{ij}^a+\varepsilon_{ij}^s$, which consists of antisymmetric (a) and symmetric (s) contributions, differences can be seen from the symmetry of the two responses. The magnon amplitude for 1.3 $\rm \mu m$ pumping does not flip sign between the two pump helicities, suggesting it arises from changes in $\varepsilon_{ij}^s$. Further, this means that linearly-polarized light can also drive the magnon, which can be explained by the inverse Cotton-Mouton effect (ICME) \cite{Kalashnikova2007}\cite{Kalashnikova2008}\cite{Tzschaschel2017}. In contrast, the sign reversal of the magnon for the two different pump helicities of the 9 $\rm \mu m$ pump originates from  $\varepsilon_{ij}^a$, consistent with the inverse Faraday effect (IFE). In this case, the circularly-polarized pulse can be treated as an effective out-of-plane magnetic field: $\overrightarrow{M}(0)\propto\overrightarrow{E}(\omega)\times \overrightarrow{E}(\omega)^*$ \cite{Kimel2005}, where $\overrightarrow{M}(0)$ is the magnetization while $\overrightarrow{E}(\omega)$ and $\overrightarrow{E}(\omega)^*$ is the electric field of the pump pulse and its complex conjugate, respectively. The amplitude and relaxation of the coherent magnon also exhibit different fluence dependences for both pump schemes, as shown in Fig. \ref{fig:4_theory}a, b. At lower fluences, the magnon oscillation amplitude A$_{magnon}$ scales linearly with the 9 $\rm \mu m$ pump fluence, while at higher fluences, it follows a quadratic scaling. On the other hand, for the 1.3 $\rm \mu m$ pump, A$_{magnon}$ increases in the low fluence regime before decreasing monotonically from 3 to 11 $\rm mJ/cm^2$ (Fig. \ref{fig:4_theory}a). In addition, the lifetime of the magnon, $\tau_{magnon}$ shows an increase (decrease) with 1.3 $\rm \mu m$ (9 $\rm \mu m$) pump fluence (Fig. \ref{fig:4_theory}b). We link the anomalous increase in lifetime for the 1.3 $\rm \mu m$ pump to noise affecting the accuracy of data extraction from the fits. Ultimately, the striking difference in the symmetry and fluence dependence suggest distinct microscopic magnon generation mechanisms for the 1.3 $\rm \mu m$ and 9 $\rm \mu m$ pumping.

We now examine the 1.3 $\rm \mu m$ (0.95 eV) pump scheme, which is resonant with charge excitations from the $J_{eff}=3/2$ band to the upper Hubbard band. A charge recovery time of $\sim2$ ps was demonstrated by previous measurements \cite{Dean2016}\cite{Hsieh2012}\cite{Afanasiev2019}. Moreover, time-resolved RIXS has shown an increase in the spectral intensity of the low-energy magnon at ($\pi$, $\pi$) with 2 $\rm \mu m$ pumping which resonantly excites electrons from $J_{eff}=1/2$ to the upper Hubbard band. This was ascribed to high-energy magnons scattering and relaxing to low-energy magnons upon charge excitations \cite{Dean2016}. We posit that a similar process is involved for 1.3 $\rm \mu m$ pumping. The charge excitation is fast and impulsive with respect to the energy of the magnon at ($\pi$, $\pi$), and magnon-magnon coupling allows higher energy magnons, i.e., at $K=(\pi$, 0) (see Methods for details), to relax to the low-energy magnon at the spin gap at ($\pi$, $\pi$). This is a complex process that involves both the charge and spin channels. A complete theoretical description is beyond the scope of this work.

For the 9 $\rm \mu m$ (138 meV) pumping, the photon energy is below the Hubbard gap, and is therefore too low to generate charge excitations, and above any optical phonon resonance (highest phonon energy being 82 meV) \cite{Moon2009} (Fig. \ref{fig:4_theory}c, d).  We note that the energy of the pump is within the two-magnon density of states, suggesting a purely magnonic excitation. A quantum theory for the inverse Faraday effect (IFE) in Sr$_2$IrO$_4$ was developed in ref. \cite{Seifert2019} which explains many aspects of the observed data. Time reversal symmetry dictates that the electric field of light couples to spin bilinears ($\sim S^iS^j$) through the spin-dependent polarization. Through coupling to the field, high-energy magnons of equal and opposite momenta are excited. Finally, via anharmonic magnon-magnon scattering, lower energy magnons are generated. This process is rapid in comparison to the magnon oscillations period and thus serves as an impulsive excitation of the precessional mode at at ($\pi$, $\pi$). The time evolution of the low-energy magnons can be modeled by equations (\ref{eq:1})-(\ref{eq:2}) presented in the Methods section.

Explicit calculation of the effective fields acting on the low-energy magnon mode from the pump (see ref. \cite{Seifert2019}) for nonresonant excitation reveals a linear dependence on the pump intensity (i.e., a quadratic dependence on the pump electric field), consistent with the experimental observations in Fig. \ref{fig:4_theory}a. A quantum treatment of the IFE is also established, explaining the sign reversal when pump helicity is reversed. The process is depicted schematically in Fig. \ref{fig:4_theory}d, which shows the magnon dispersion (for details, see Methods) around ($\pi, \pi)$. The coupling of the mid-IR pump pulse to the low-energy $B_{2g}$ magnon involves three steps. First, the 9 $\rm \mu m$ (138 meV) pump pulse directly couples to virtual pairs of high-energy magnons with equal and opposite momenta through the spin-dependent electric polarization. Subsequently, the presence of the pump pulse generates anharmonic magnon-magnon interactions, which allows the high-energy magnons to scatter and relax down to the gap at ($\pi$, $\pi$). Finally, since both the pump and magnons are high-energy compared to the gap, these fast interactions act as an effective impulsive field on the in-plane $B_{2g}$ magnon, driving the coherent magnon mode detected by MOKE. Taking both fluence dependence and penetration depth into account, we showed above that 9 $\rm \mu m$ pulses generate magnons with an energy density two orders of magnitude lower than 1.3 $\rm \mu m$ pulses. Consequently, direct magnon excitation with below-gap 9 $\rm \mu m$ driving results in a magnon generation efficiency two orders of magnitude higher than that of above-gap excitation, enabling extremely efficient ultrafast manipulation of magnetism.

\section*{Outlook}
Our work unravels a previously-unexplored mechanism to control and generate magnons in the spin-orbit coupled antiferromagnet Sr$_2$IrO$_4$ using ultrafast laser pulses. The key result is that direct excitation of the spin spectrum leads to nonlinear magnon generation with two orders of magnitude higher efficiency compared to excitation that is resonant with charge excitations. This is consistent with the mechanism proposed in ref. \cite{Seifert2019} that involves mid-IR light-magnon coupling through generation of high-energy magnons that scatter and relax to the low-energy $B_{2g}$ magnon at ($\pi$, $\pi$) being mediated by anharmonic magnon-magnon interaction. We note that photoexcitation at other pump energies (within the magnon bandwidth) also holds possibilities to enhance magnon generation efficiency since the effective fields are strongest when the pump is tuned to the maximum of the two-magnon density of states, which in this case, is at the edge of the Brillouin zone. Moreover, these results have broader implications for below-gap driving coherent magnon generation which can be applied to a suite of exotic magnetic materials. Overall, our observation of the coherent magnon establishes time-resolved MOKE as a powerful probe of magnetoelastic effects in quantum magnets, which can include 2D van der Waals materials with unusual magnetic properties. Our results may unlock exciting opportunities in spintronics by efficient nonlinear magnon generation, and may motivate research in spin qubit-based quantum computing. Finally, these findings could positively contribute to the study of topologically nontrivial magnon excitations which may contain Berry curvature information as well as nonzero Chern numbers.

\newpage
\section*{Methods}
\subsection*{Sample growth}
Millimeter-scale single crystals of Sr$_2$IrO$_4$ were grown using a self-flux technique from off-stoichiometric quantities of IrO$_2$, SrCO$_3$ and SrCl$_2$. The ground mixtures of powders were melted at 1,470$\degree$C in partially-closed platinum crucibles. After a soaking phase of more than 20 hours, the mixture was slow-cooled at 2°C$\rm h^{-1}$ to reach 1,400$\degree$C. Then the crucible was cooled down to room temperature at a rate of 100°C$\rm h^{-1}$.

\subsection*{Experimental setup}
Extended Data Fig. \ref{extfig:1_schematic} shows a schematic of our experiment. We used a Spitfire Ace regenerative amplified titanium:sapphire laser amplifier system with pulse energy of 6 mJ, pulse duration of 100 fs, 1 kHz repetition rate and 800 nm central wavelength as the main light source. The incoming beam from the light source is split into two beams using a beam splitter: one beam is used for generating pump pulses while the second beam is used to generate the probe pulses. 

The first beam is further divided in two paths via another beam splitter, and the resultant two 800 nm beams have 1.9 mJ energy each. These beams are injected into a TOPAS-TWINS system which consists of two 2-stage optical parametric amplifiers (OPA). Each OPA is tunable, and generates signal beams with wavelengths from 1.16 - 1.6 $\rm \mu m$ and idler beams from 1.6 - 2.6 $\rm \mu m$. 

The 9 $\rm \mu m$ mid-IR beam with stable carrier envelope phase (CEP) is generated by mixing two near-IR signal beams (tunable from 1.1 - 2 $\rm \mu m$) from the output of the TOPAS-TWINS in a 0.6 mm thick GaSe crystal. The generation mechanism is type-II difference frequency generation (DFG) where the mid-IR frequency is determined by the frequency difference and phase matching condition of the two near-IR beams mixing in the GaSe crystal. To accomplish this, one output from the TOPAS-TWINS was fixed at 1.2 $\rm \mu m$ while the other output was tunable from 1.2 - 1.5 $\rm \mu m$. A delay stage stage was used to control the delay between these two beams. 

The 1.3 $\rm \mu m$ pump pulses are directly from one of the outputs of the TOPAS-TWINS. In order to generate circularly-polarized pump beams, a quarter-wave plate was placed after the GaSe crystal that enabled light ellipticity control. The incidence angles for pump and probe pulses on the sample are around 20° and 10°, respectively. 

The 800 nm  probe pulse is reflected off the sample, and propagates through a half-wave plate and a Wollaston prism which divides the signal into vertically- and horizontally-polarized components. The two components then reach two balanced photodiodes. The pump-induced polarization rotation can be measured by the difference of photocurrents $\Delta I$ between the two photodiodes, and the MOKE signal $\Delta I/2I$ (2I is the sum of the photocurrents) is recorded as a function of the time delay between the pump and probe pulses. For signal detection, a Stanford Research System lock-in amplifier (Model SR830) was utilized which was triggered by the output from a phase-locked optical chopper (chooped at 500 Hz frequency, model: Newport/New Focus 3502) used to modulate the pump pulse. 

In order to perform cryogenic measurements, the sample was placed inside a Janis ST-300 cryostat connected to a liquid He dewar, under vacuum condition. This allowed for low-temperature measurements down to 9 K.

\subsection*{Peak electric field calculation}
For a given fluence $F$, the peak electric field of the mid-IR beam can be estimated from the formula, $E_{0}=\sqrt{\frac{F}{2 \varepsilon_0 c \Delta t}}$. Here, $c$ is the speed of light, $\varepsilon_0$ is the vacuum permittivity, and $\Delta t$ is the pulse duration (100 fs in this experiment). The maximum pump pulse energy used in the measurement was 26 nJ, which was focused down to a spot size of 60 $\rm \mu m$. This corresponds to a maximum fluence of 0.92 $\rm mJ/cm^2$, attaining a peak electric field of 1.4 MV/cm.

\subsection*{Penetration depth calculation}
The penetration depth of light $\delta$  is calculated based on the real part of the optical conductivity $\sigma_1$ and real part of the permittivity $\varepsilon_1$ \cite{Propper2016}. It can be written as $\delta=\lambda/ 2\pi k$, where, $k= \rm Im(\sqrt{\varepsilon_1+ i \varepsilon_2})$. In this relation, $\varepsilon_2$ is the imaginary part of the permittivity that is obtained from $\sigma_1$.

We have computed the penetration depth of 800 nm probe, 1.3 $\rm \mu m$ pump and 9 $\rm \mu m$ to be 157 nm, 141 nm and 1217 nm, respectively. It is noted that the 1.3 $\rm \mu m$ pump has similar penetration depth as the probe, while the 9 $\rm \mu m$ pump penetrates nearly one order of magnitude deeper into the sample compared to the probe.

\subsection*{Pump-induced heating}
Heating effects must be considered in all pump probe measurements since the effective lattice temperature after the intensive pump is different than the equilibrium temperature. We use the two-temperature model to estimate the effective lattice temperature after the electron-phonon thermalization (i.e. a few picoseconds). The specific heat of Sr$_2$IrO$_4$ is described as: $C_{p}(T)=\gamma T^2+\beta T^3$, with $\gamma = 2$ $\rm mJ mol^{-1} K^{-2}$, $\rm \beta = 0.5 $ $\rm mJ mol^{-1} K^{-3}$. For a given pump fluence we estimate the absorbed energy on the photo-excited area and calculate the effective temperature after the electron-phonon thermalization by integrating the equation: 
\begin{equation}
    Q_{pump}=\int_{T_i}^{T_f} N C_p(T) \,dT.
\label{eq:heat}
\end{equation}

Here, $T_i$ and $T_f$ are the initial and final temperatures, $N$ is the number of moles in the excited volume. The absorbed energy $Q_{pump}$ is estimated by the relation, $Q_{pump} = F A (1-R)$, where $F$ is the fluence, $A$ is the beam size on the sample, and $R$ is the reflectivity. The photoexcited volume of the sample is estimated using the pump penetration depth $\delta$ and the beam size \cite{Haque2024}.

Using these relations, we calculated and plotted the effective temperature upon the 9 $\rm \mu m$ and 1.3 $\rm \mu m$ pump excitations with initial temperature of 10 K and 80 K, respectively in Extended Data Fig. \ref{extfig:2_heat}. With the highest fluence we used for 9 $\rm \mu m$ pump (0.92 $\rm mJ/ cm^{2}$) and 1.3 $\rm \mu m$ pump (11.5  $\rm mJ/ cm^{2}$), it was observed that the effective temperature was well below the Néel temperature, indicating the non-thermal nature of the coherent magnon control.

\subsection*{Spin spectrum of Sr$_2$IrO$_4$}
The spin spectrum of Sr$_2$IrO$_4$ around ($\pi$, $\pi$) in Fig. \ref{fig:4_theory}d is calculated using SpinW library \cite{Toth2015}. The magnetic interaction Hamiltonian for ($\pi$, $\pi$) is: $H=H_{H}+H_{DM}+H_{spin-lattice}$ \cite{Porras2019}\cite{Liu2019}. 

The Heisenberg term is $H_H=\sum_{\left \langle i,j \right \rangle}J_{ij} \overrightarrow{S_i} \cdot \overrightarrow{S_j}$, where $\overrightarrow{S_i}$ is the pseudospin at site $i$ and $J_{ij}$ corresponds to in-plane first, second, third nearest-neighbor interactions $J_1$, $J_2$, $J_3$ and next first and second interlayer nearest-neighbor interactions $J_{1l}$, $J_{2l}$. The Dzyaloshinsky-Moriya (DM) interaction Hamiltonian is written as, $H_{DM}=\sum_{\left \langle i,j \right \rangle} J_{z}S_i^zS_j^z + \overrightarrow{D} \cdot (\overrightarrow{S_j} \times \overrightarrow{S_j})$. Here, $\overrightarrow{D}$ is the out-of-plane DM vector, inducing the spin canting of 13\degree. The final term is the lattice distortion-induced exchange anisotropy: $ H_{spin-lattice}=\sum_{\left \langle i,j \right \rangle}\Gamma(S_i^xS_j^y+S_i^yS_j^x) $, where $\Gamma$ is the anisotropy strength.

Extended Data Fig. \ref{extfig:3_spin_spec} depicts the full in-plane spin spectrum along a-axis across the full Brillouin zone based on the linear spin wave theory. The 2 meV spin gap that we observed is located at ($\pi$, $\pi$). Thus, we limit our interest to the spin excitations around ($\pi$, $\pi$).

\subsection*{Effective fields acting on low-energy magnons upon photoexcitation}
After photoexcitation with 9 $\rm \mu m$ mid-IR circularly-polarized pump, the electric field of the light couples to the spin bilinears and generates anharmonic magnon-magnon interactions. The mid-IR pump pulse is a high-energy degree of freedom, compared to the spin gap. As such, along with the high-energy magnons, these high-energy pump modes can be integrated out to obtain effective magnetic fields acting on the low-energy magnons. The generalized dynamics is governed by the following equations of motion,
\begin{equation}
    \partial_t u =\chi^{-1}m-\frac{u}{\tau_u}-h_m,
\label{eq:1}
\end{equation}
\begin{equation}
    \partial_t m = -\kappa u -\frac{m}{\tau_m}+h_u.
\label{eq:2}  
\end{equation}

Here, $m$ is the out-of-plane magnetization, $u$ is a linearized angle for the deviation of the in-plane Ne\'el AFM order parameter from its equilibrium ordering axis, $\chi^{-1}$ is the out-of-plane susceptibility, $\kappa$ is proportional to an in-plane anisotropy. In addition, $\tau_{m,u}$ and $h_{m,u}$ are the phenomenological relaxation times and light-induced effective fields, respectively. 

In the ultrafast regime, when the pump pulse is short, the effective field $h_m$ acts as an impulse providing an initial velocity, while $h_u$ provides an initial amplitude of $m$. To prove this, we first write write the equation of motion for $m(t)$, which is given by
\begin{equation}
    \partial_t^2 m + 2 \gamma \partial_t m + \omega_0^2 m= \kappa h_m + \partial_t h_u.
\label{eq:4}
\end{equation}

This resembles a driven harmonic oscillator where $ \gamma=\frac{1}{2} (\frac{1}{\tau_m}+\frac{1}{\tau_u})$ and $\omega_0^2=\kappa/\chi$. The trivial homogeneous solution $m=0$ is derived for initial conditions $m(0)=\partial_t m(0)=0$. For the inhomogeneous solution, we use the Green's function for a harmonic insulator,
\begin{equation}
    G(t-t')= \int_{-\infty}^{\infty} \frac{d\omega}{2\pi} \frac{e^{-i \omega (t-t')}}{\omega^2+2 i \gamma \omega-\omega_0^2}.
\label{eq:5}
\end{equation}

In the next stage, the magnetization dynamics $m_m (t)$ and $m_u (t)$ on the presence of effective fields $h_m$ and $h_u$, respectively, are obtained by the convolution:
\begin{equation}
    m_m (t)= \int_{-\infty}^{\infty} dt' h_m(t') G(t-t'),
\label{eq:6}
\end{equation}
\begin{equation}
    m_u (t)= -\int_{-\infty}^{\infty} dt' h_u(t') \partial_{t'}G(t-t'),
\label{eq:7}
\end{equation}
where we used $h_u(\pm \infty)=0$. Taking causality into account, unit pulses of strength $\Bar{h}$ and duration $t_p$ starting at $t=0$ can be written as $h_{m,u}(t)=\Bar{h}_{m,u}[\Theta(t)-\Theta(t-t_p)]$. Plugging this into (\ref{eq:6}) and (\ref{eq:7}) yields
\begin{multline}
    m_m (t)= \frac{\kappa \Bar{h}_m}{\Bar{\omega} (\gamma^2+\Bar{\omega}^2)} [\Theta(t) [\Bar{\omega}-e^{-\gamma t}(\gamma \sin \Bar{\omega}t + \Bar{\omega} \cos \Bar{\omega}t )]\\
    -\Theta(t-t_p)[\Bar{\omega}-e^{-\gamma (t-t_p)}(\gamma \sin \Bar{\omega}(t-t_p) + \Bar{\omega} \cos \Bar{\omega}(t-t_p))]],
   \label{eq:8} 
\end{multline}
\begin{equation}
    m_u (t)= \frac{\Bar{h}_u}{\Bar{\omega}} [\Theta(t) e^{-\gamma t} \sin \Bar{\omega}t
    -\Theta(t-t_p) e^{-\gamma (t-t_p)} \sin \Bar{\omega}(t-t_p)],
   \label{eq:9} 
\end{equation}
where $\Bar{\omega}=\sqrt{\omega_0^2-\gamma^2}$ is the damped eigenfrequency. When the pump pulse is turned off at $t=t_p$, the magnetization $m(t)$ temporally evolves according to the homogeneous equations of motion with initial conditions set by the $m_{m,u}(t)$ shortly after $t_p$. We evaluate $m_{m}(t)$ and its derivative at $t=t_p^+$ which gives
\begin{equation}
    m_m (t_p^+)= \frac{\kappa \Bar{h}_m}{\Bar{\omega} (\gamma^2+\Bar{\omega}^2)} e^{-\gamma t_p}[\Bar{\omega}e^{\gamma t_p}-\gamma \sin \Bar{\omega}t_p - \Bar{\omega} \cos \Bar{\omega}t_p ]=\mathcal{O}(t_p^2),
   \label{eq:10} 
\end{equation}
\begin{equation}
    \partial_t m_m (t_p^+)= \frac{\kappa \Bar{h}_m}{\Bar{\omega}} e^{-\gamma t_p}\sin \Bar{\omega}t_p =\kappa \Bar{h}_m t_p + \mathcal{O}(t_p^2).
   \label{eq:11} 
\end{equation}

Following the similar procedure for $m_u(t)$, we obtain
\begin{equation}
    m_u (t_p^+)= \frac{\Bar{h}_u}{\Bar{\omega}} e^{-\gamma t_p}\sin \Bar{\omega}t_p =\Bar{h}_u t_p + \mathcal{O}(t_p^2),
   \label{eq:12} 
\end{equation}
\begin{equation}
    \partial_t m_u (t_p^+)= -\frac{\Bar{h}_u}{\Bar{\omega}} e^{-\gamma t_p}[\Bar{\omega}e^{\gamma t_p}+\gamma \sin \Bar{\omega}t_p - \Bar{\omega} \cos \Bar{\omega}t_p ]=\mathcal{O}(t_p^2).
   \label{eq:13} 
\end{equation}

It is evident from equations (\ref{eq:10})-(\ref{eq:11}) that for a short pulse $t_p\ll \Bar{\omega}^{-1}, \gamma^{-1}$, the effective field $h_m$ acts as an impulsive drive by giving the magnetization a finite initial velocity $\partial_t m$. On the other hand, equations (\ref{eq:12})-(\ref{eq:13}) show that $h_u$ provides an initial amplitude for the oscillations by displacing the magnetization.

According to the calculations, $h_{u}$ and $h_{m}$ contain terms which are maximized for circularly-polarized excitations, and flip sign for opposite helicity. In addition, other terms are maximized for linearly-polarized excitation. The strengths of all these terms depend on the pump frequency through the two-magnon density of states as well as coupling of the pump pulse electric field to different spin bilinears (see ref. \cite{Seifert2019} for more details). As a consequence, this formalism provides a fresh perspective on distinct magnon excitation mechanisms as a function of pump energy and polarization. Importantly, the magnetization $m$ and (linearized) angle $u$ are coupled. As such, driving either $m$ or $u$ can induce the magnetization dynamics which can be tracked by the MOKE probe (Extended Data Fig. \ref{extfig:4_protocol}a-b).


\section*{Data availability}
The data presented in this manuscript can be available from the corresponding authors upon reasonable request. Correspondence should be addressed to Sheikh Rubaiat Ul Haque or Richard D. Averitt.

\section*{Acknowledgements}
We thank Gil Refael, David Hsieh, Dmitri N. Basov, Nuh Gedik, Andrea F. Young, Susanne Stemmer, Tony F. Heinz, Aaron M. Lindenberg, Shengxi Huang, and Daniel P. Arovas for fruitful discussions. This work is supported by ARO MURI grant no. W911NF-16-1-0361. Part of the data analysis was carried out at SLAC and supported by the U.S. Department of Energy (DOE), Office of Science, Office of Basic Energy Sciences (BES), Materials Sciences and Engineering Division.

\section*{Author contribution}
R.D.A. conceived the project. X.C. performed the material growth and characterization of the sample under the supervision of S.D.W. G.-F.Z., J.Z. and K.A.C. built the experiment. G.-F.Z., S.R.U.H. and K.J.K. conducted the experiment and analyzed the data. The spin spectrum was calculated by S.R.U.H. U.F.P.S. performed the theoretical calculations under the guidance of L.B. R.D.A. and L.B. supervised the project. G.-F.Z., S.R.U.H., K.J.K. and R.D.A. wrote the manuscript with input from all the authors. 

\section*{Competing interests}
The authors declare no competing interests.

\newpage
\setcounter{figure}{0}
\renewcommand{\figurename}{\textbf{Extended Data Fig.}}
\renewcommand{\thefigure}{\arabic{figure}}
\begin{figure} [hbt!]
    \centering
    \includegraphics[width=150 mm]{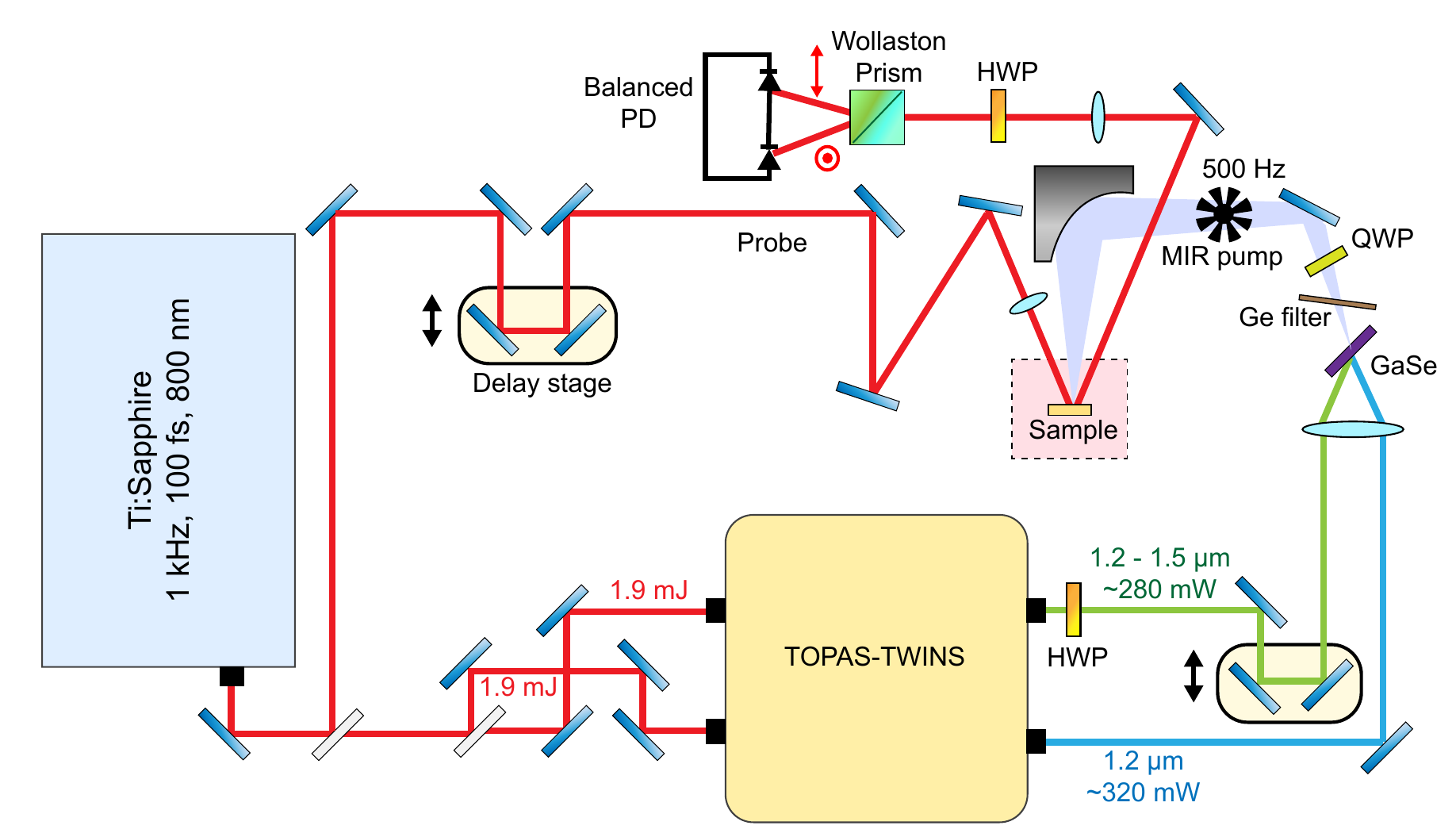}
    \caption{\textbf{Schematic of the experimental configuration.} Experimental setup for  9 $\rm \mu m$ or 1.3 $\rm \mu m$ pump - 800 nm magneto-optic Kerr effect (MOKE) probe spectroscopy. HWP: half-wave plate, QWP: quarter-wave plate, MIR: mid-IR, PD: photodiode. The 9 $\rm \mu m$ mid-IR beam is generated by mixing two near-infrared beams (tunable from 1.1 to  2 $\rm \mu m$) from the output of a TOPAS-TWINS dual OPA system in a GaSe crystal while the 1.3 $\rm \mu m$ pump pulses are directly from one of the outputs of the TOPAS-TWINS. }
    \label{extfig:1_schematic}
\end{figure}

\newpage
\begin{figure} [hbt!]
    \centering
    \includegraphics[width=140 mm]{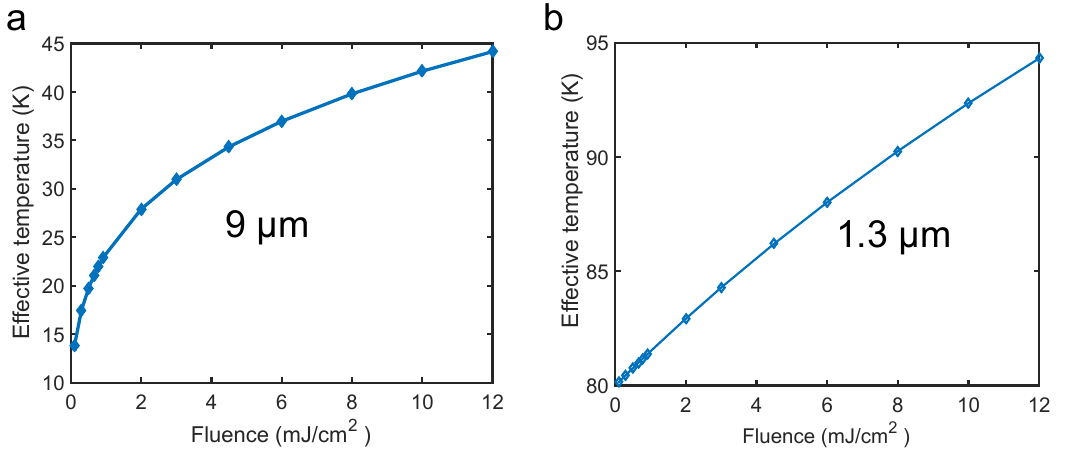}
    \caption{\textbf{Pump-induced heat}. The effective temperature of the sample based on the two-temperature model upon 9 $\rm \mu m$ pumping at 10 K (\textbf{a}) and 1.3 $\rm \mu m$ pumping at 80 K (\textbf{b}). }
    \label{extfig:2_heat}
\end{figure}

\newpage
\begin{figure} [hbt!]
    \centering
    \includegraphics[width=130 mm]{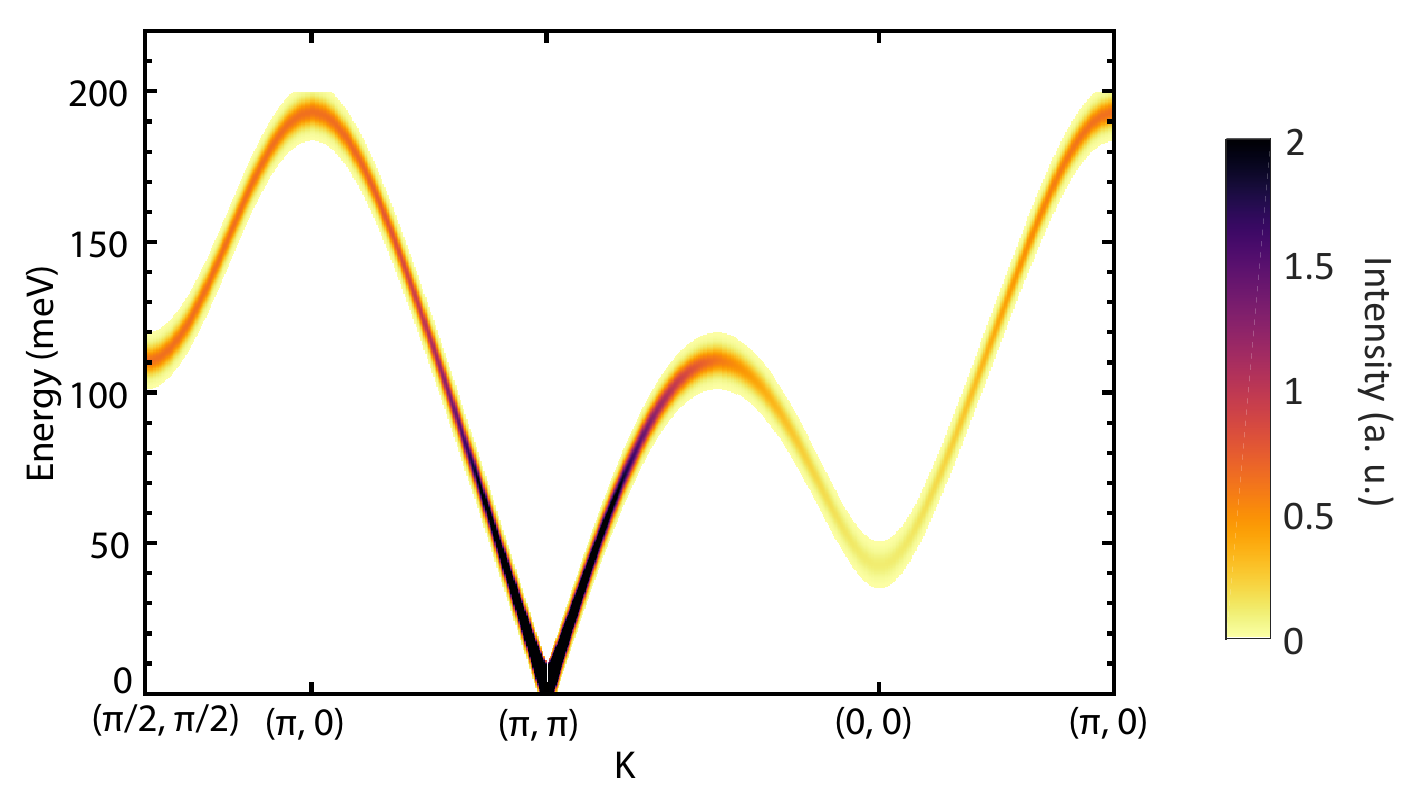}
    \caption{\textbf{Full spin spectrum of Sr$_2$IrO$_4$}. The parameters used for simulation are the same as in ref. \cite{Porras2019}: $J_1 = 57$ meV, $J_2 = -16.5$ meV, $J_3= 12.4$ meV, $J_{1l} = 16.4$ $\rm \mu eV$, $J_{2l} = -6.2$ $\rm \mu eV$, $J_z = 2.9$ meV, $D = 28$ meV, and $\Gamma = 2.7$ $\rm \mu eV$. Gaussian broadening $\delta E = 7$ meV is used for clarity. }
    \label{extfig:3_spin_spec}
\end{figure}

\newpage
\begin{figure} [hbt!]
    \centering
    \includegraphics[width=120 mm]{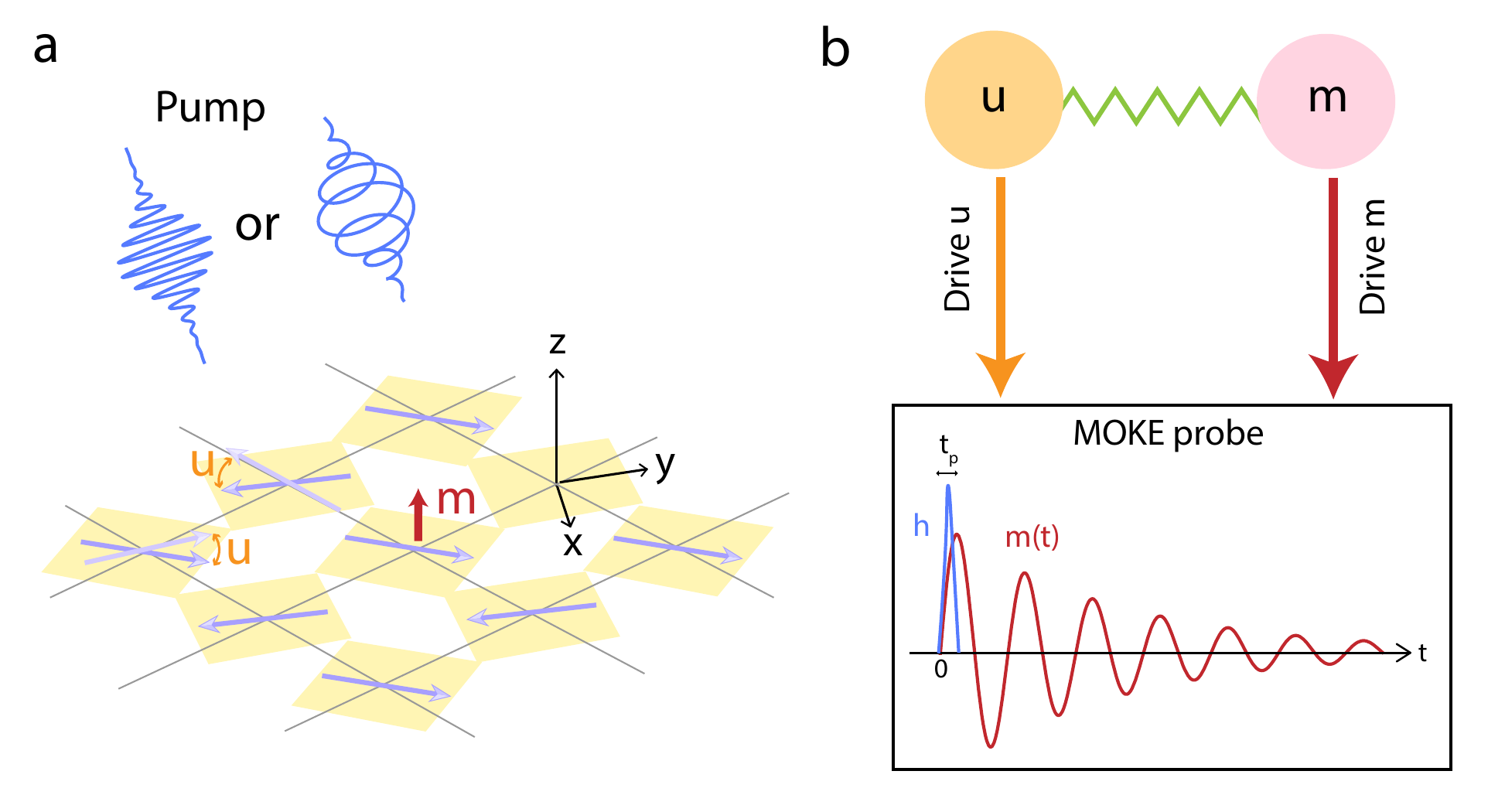}
    \caption{\textbf{Microscopic mechanism of MOKE in Sr$_2$IrO$_4$. a}, Circularly- or linearly-polarized pump exciting the sample by driving small angle fluctuation from the AFM ordering axis $u$ or out-of-plane magnetization $m$. The MOKE probe captures the photoinduced dynamics by monitoring the Kerr rotation angle which is proportional to the magnetization. \textbf{b}, Green zigzag line shows coupling between $m$ and $u$. As such, driving either $m$ or $u$ can produce oscillations in magnetization $m$. For a short pump pulse, its electric field induces an effective field $h$ (blue curve) which acts as an impulsive drive and sets initial condition for the dynamics of $m$, manifested in a magnon oscillation detercted by the MOKE probe (red curve).}
    \label{extfig:4_protocol}
\end{figure}

\end{document}